\documentclass[sigconf]{acmart}
\usepackage[font=small]{caption}
\usepackage[labelformat=simple]{subcaption}
\usepackage[ruled,vlined,linesnumbered]{algorithm2e}
\usepackage{algorithmic}
\SetKwComment{Comment}{$\triangleright$\ }{}
\usepackage[mathscr]{eucal}
\usepackage{color}
\usepackage{url}
\usepackage{mathtools}
\usepackage{xcolor}
\usepackage{multirow}
\usepackage{soul}
\usepackage{colortbl}
\usepackage{balance}

\renewcommand{\textrightarrow}{$\rightarrow$}

\newcommand{\method}{\textsc{CASPER}\xspace}

\newcommand{\ltcross}{\textsc{LatentCross}\xspace}
\newcommand{\tisasrec}{\textsc{TiSASRec}\xspace}
\newcommand{\lstm}{\textsc{LSTM}\xspace}

\newcommand{\jodie}{\textsc{JODIE}\xspace}

\newcommand{\argmax}[1]{\underset{#1}{\operatorname{arg}\,\operatorname{max}}\;}

\copyrightyear{2022}
\acmYear{2022}
\setcopyright{acmcopyright}\acmConference[CIKM '22]{Proceedings of the 31st ACM
International Conference on Information and Knowledge Management}{October
17--21, 2022}{Atlanta, GA, USA}
\acmBooktitle{Proceedings of the 31st ACM International Conference on Information
and Knowledge Management (CIKM '22), October 17--21, 2022, Atlanta, GA, USA}
\acmPrice{15.00}
\acmDOI{10.1145/3511808.3557425}
\acmISBN{978-1-4503-9236-5/22/10}

\begin{document}

%%
%% The "title" command has an optional parameter,
%% allowing the author to define a "short title" to be used in page headers.
\title{Rank List Sensitivity of Recommender Systems to\\ Interaction Perturbations}

%%
%% The "author" command and its associated commands are used to define
%% the authors and their affiliations.
%% Of note is the shared affiliation of the first two authors, and the
%% "authornote" and "authornotemark" commands
%% used to denote shared contribution to the research.
\author{Sejoon Oh}
\email{soh337@gatech.edu} 
\affiliation{%
  \institution{Georgia Institute of Technology}
    \country{United States}
}

\author{Berk Ustun}
\email{berk@ucsd.edu}
\affiliation{%
  \institution{University of California San Diego}
    \country{United States}
}

\author{Julian McAuley}
\email{jmcauley@eng.ucsd.edu} 
\affiliation{%
  \institution{University of California San Diego}
    \country{United States}
}

\author{Srijan Kumar}
\email{srijan@gatech.edu} 
\affiliation{%
  \institution{Georgia Institute of Technology}
    \country{United States}
}

%%
%% By default, the full list of authors will be used in the page
%% headers. Often, this list is too long, and will overlap
%% other information printed in the page headers. This command allows
%% the author to define a more concise list
%% of authors' names for this purpose.
\renewcommand{\shortauthors}{Oh, et al.}

	\begin{abstract}
		\label{sec:abstract}
		Prediction models can exhibit sensitivity with respect to training data:  small changes in the training data can produce models that assign conflicting predictions to individual data points during test time.
In this work, we study this sensitivity in recommender systems, where users' recommendations are drastically altered by minor perturbations in other unrelated users' interactions. 
We introduce a measure of stability for recommender systems, called \textit{Rank List Sensitivity} (RLS), which measures how rank lists generated by a given recommender system at test time change as a result of a perturbation in the training data.
We develop a method, \method, which uses cascading effect to identify the minimal and systematical perturbation to induce higher instability in a recommender system. 
Experiments on four datasets show that recommender models are overly sensitive to minor perturbations introduced randomly or via \method — even perturbing one random interaction of one user drastically changes the recommendation lists of all users. 
Importantly, with \method perturbation, the models generate more unstable recommendations
for low-accuracy users (i.e., those who receive low-quality recommendations) than high-accuracy ones.
	\end{abstract}

%%
%% This command processes the author and affiliation and title
%% information and builds the first part of the formatted document.
\maketitle
	
	\section{\textbf{Introduction}}
	\label{sec:intro}
	
\begin{figure}[t]
    \centering
    \includegraphics[width=8.0cm]{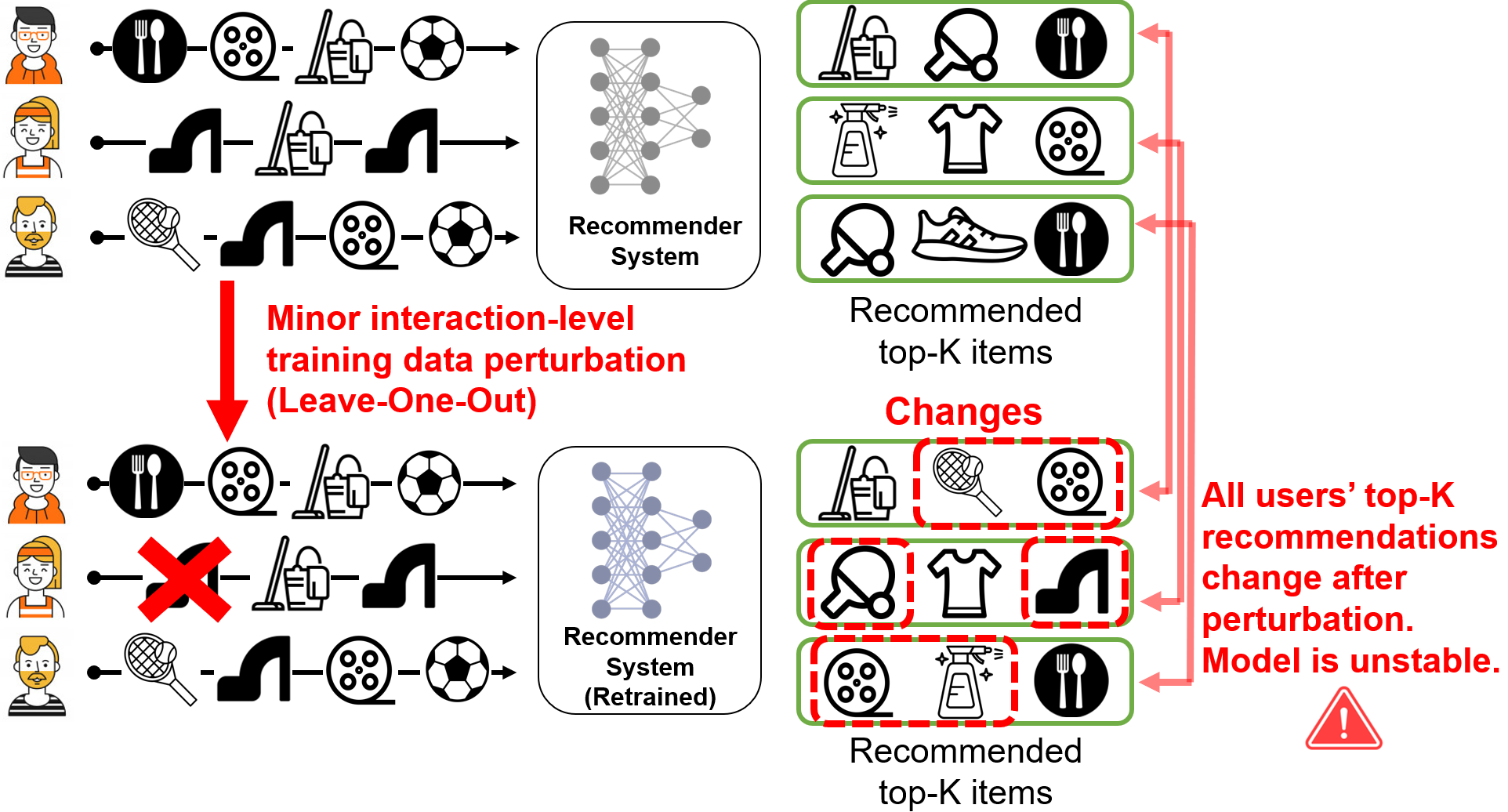}
    \caption{
    Small changes (e.g., leave-one-out perturbation) in the training data can produce recommender systems that output drastically different recommendations for \textit{all} individual users.
    }
    \label{fig:main_figure}
\end{figure}

Small changes in training data can produce large changes in outputs of machine learning models~\cite{steinhardt2017certified, yang2017generative, universal_attack, sarkar2017upset, NIPS2016_6142, RL_attack}.
\citet{marx2020predictive} showed how classification tasks can often admit competing models that perform almost equally well in terms of an aggregate performance metric (e.g., error rate, AUC) but that assign conflicting predictions to individual data points. 
Likewise, \citet{black2021leave} showed how removing a point from a training dataset can produce models that assign drastically different predictions, and highlighted how this lack of stability disproportionately affects points with low confidence predictions.

The sensitivity with respect to minor data perturbations is especially meaningful and concerning in modern recommender systems -- where data points pertain to user interactions. 
In this setting, sensitivity would imply that the recommendations for a user change due to small arbitrary changes in the training data from another unrelated user. This effect can be disruptive or even dangerous when recommendation systems are used for applications in healthcare, finance, education, and housing~\cite{tran2021recommender,zibriczky122016recommender, tan2008learning, yuan2013toward}. Consider a system that recommends a specific treatment to a patient based on data from their electronic health record~\cite{tran2021recommender, wiesner2014health}.  In this setting, sensitivity would imply that the treatment recommendations for a patient by a given system could change due to noisy training data for another patient -- e.g., due to errors introduced when digitizing hand-written notes or transcribing voice memos~\cite{cruz2009data, sun2018data, kumah2018electronic}. 
More broadly, this sensitivity could be introduced due to intentional manipulations -- as a malicious adversary could inject noise into the training data to degrade the overall recommendation quality by producing low-quality recommendations for all users 
or even disproportionate damage on specific user or item groups~\cite{zhang2020practical, fang2020influence, PoisonRec, fang2018poisoning,taamr}. 

These effects broadly underscore the need to \textit{measure} the sensitivity in the development of recommender systems -- so that model developers and end users can decide whether to use a specific recommendation, or whether to use recommender systems at all. 

In this work, we study the sensitivity of recommender systems, so that practitioners can measure the stability of their models and make informed decisions in model development and deployment. Our problem statement is: 
\textbf{\textit{Can an arbitrary change in a single data point in the training data change the recommendations for other data points? If so, what is the maximum change in recommendations possible with that change?}}

We propose a novel framework to measure the stability by comparing two recommendation lists for each test interaction -- the recommendation list from a recommender model trained on the original training data, and the recommendation list from a model trained on the perturbed training data. Then, the two recommendation lists are compared for each test interaction as shown in Figure \ref{fig:main_figure}. If the two lists are the same, then we say that the model is stable to the perturbation; otherwise, the model is unstable.  

Our approach requires a metric to differentiate the order of items between two lists. Standard next-item prediction metrics such as MRR, Recall, NDCG, and AUC are applicable to one list (by measuring the rank of the ground-truth next item in the list). 
Extensions of these metrics, e.g., the difference in MRR, are not appropriate since those metrics can remain unchanged even if the rank list is drastically different, but the ground-truth item's rank remains similar. This happens in practice (see Figure~\ref{tab:next_item_metrics}). 
Thus, we introduce a formal metric to quantify the stability of recommender systems, namely \textit{Rank List Sensitivity} (RLS), which measures the similarity in the rank lists generated in the presence versus absence of perturbations. 
We employ two metrics to measure RLS, namely Rank-Biased Overlap (RBO)~\cite{RBO} and Jaccard Similarity~\cite{jaccard1912distribution}. RBO measures similarity in the order of items in two lists, while the Jaccard score highlights the overlap in the top-K items without considering their order. Higher scores in both metrics are better. 

\begin{figure}[t!]
\hspace{-5mm}
\begin{subfigure}{0.23\textwidth}
    \centering
    \includegraphics[width=4.6cm]{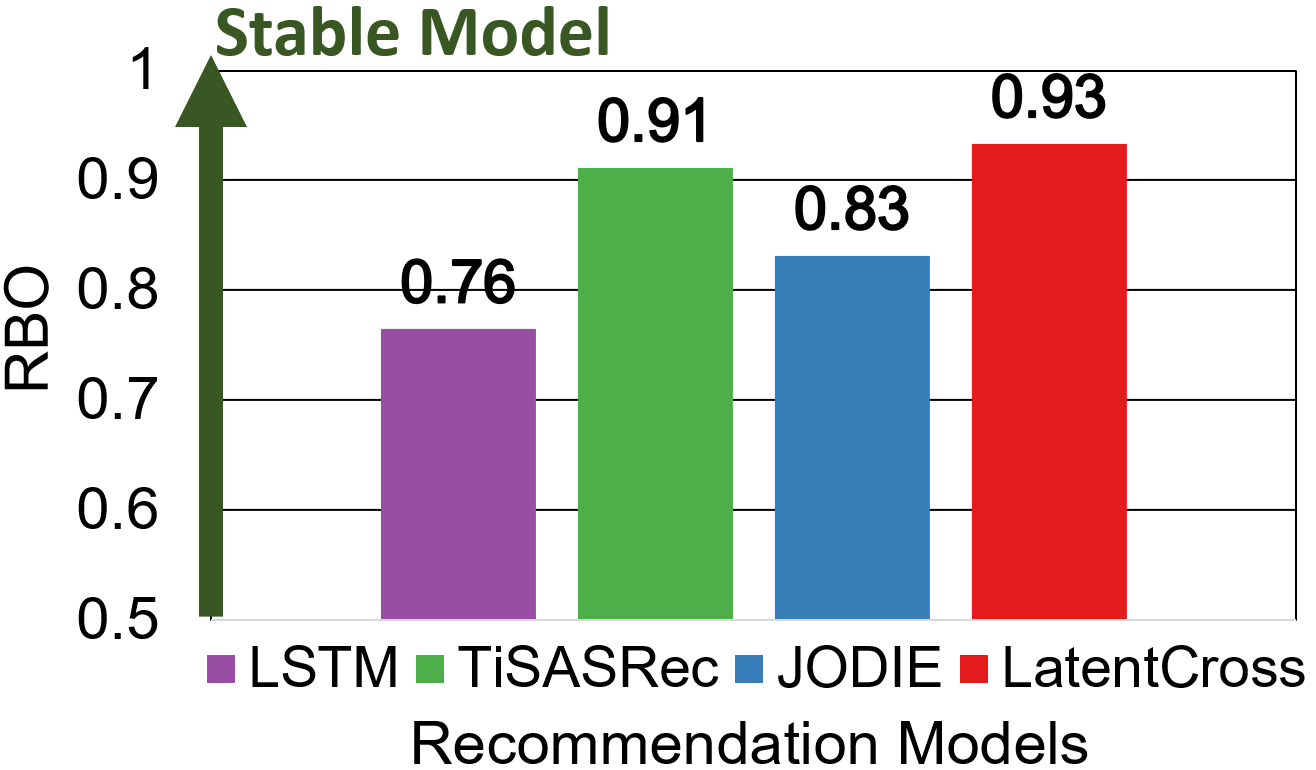}
    \caption{Random LOO perturbation}
    \label{fig:random_deletion_intro}
    \end{subfigure}
    \hfill 
    \begin{subfigure}{0.23\textwidth}
    \centering
    \includegraphics[width=4.2cm]{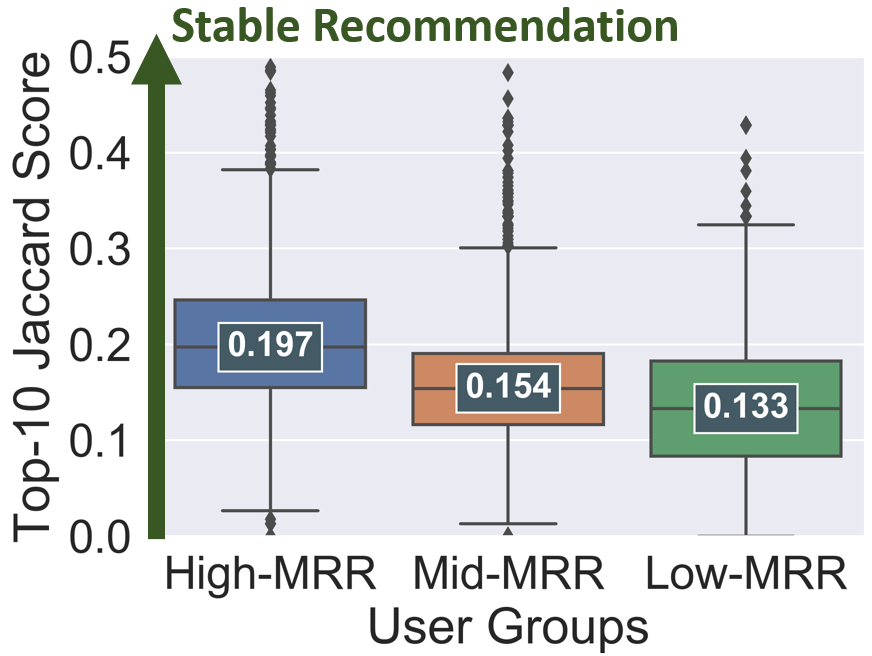}
    \caption{Stability across user groups}
    \label{fig:user_analysis}
    \end{subfigure}
    \caption{ (a) Four recommendation models are shown to be unstable against minor perturbations --- random leave-one-out perturbation in training data changes output rank lists of \textit{all users} drastically. 
    (b) Our proposed perturbation method, \method, lowers the stability, measured via Jaccard@10, of users with low accuracy the most.}
\end{figure}

We introduce two training data perturbation methods to measure the stability of recommender systems: random perturbations and \method perturbations. 
Random perturbations select one interaction out of all training data interactions randomly for perturbations. Using random perturbations, we can measure the model sensitivity caused due to arbitrary errors and noise. 
On the other hand, \method\ is designed to identify an interaction whose perturbation can introduce higher instability in recommendations than random perturbations. 
Such interaction reveals model vulnerabilities that can potentially be exploited by adversaries to manipulate the recommender system. 
To find the deliberate perturbation,
we hypothesize a cascading effect by creating an interaction-to-interaction dependency graph. 
Then, \method perturbs an interaction with the largest number of descendants in the graph, which leads to significant changes in the generated recommendations. 
\method is fast and scalable to the dataset size and does not require model parameters or gradients to identify the perturbation.

Experimentally, we first investigate the sensitivity of models to random perturbations. 
We show that the recommender models are sensitive to random interaction perturbations. 
Even \textit{one} random interaction perturbation drastically changes the entire rank lists of items for \textit{all} users. This is shown as low RBO scores (lower than 1.0 score means the rank list has changed) of four recommendation models on Foursquare (Figure~\ref{fig:random_deletion_intro}) and all four datasets (shown later in Figure~\ref{fig:random_perturbation}), and as low top-10 Jaccard scores (Figure~\ref{fig:top10_jaccard}). We underline that the instability of the models occurs due to the data change, not the training randomness (e.g., different random seed, initialization, etc.), since we remove all the randomness during the training to focus solely on the effect of training data perturbation.

Next, we compare \method\ with five training data perturbation algorithms.
We show that \method\ identifies a perturbation to be made that reveals higher sensitivity in recommendation models compared to existing methods across datasets.
Importantly, we find that \method\ identifies an interaction whose perturbation results in low-accuracy user groups being more impacted as per model sensitivity --- the top-10 Jaccard scores are lower for low-MRR users than for high-MRR users (see Figure~\ref{fig:user_analysis}).
Since the item ranking in the recommendation list has a significant impact on user satisfaction~\cite{pei2019personalized}, if the recommendation is low-quality and unstable, the user satisfaction and engagement can be dramatically reduced, and it may result in user dropout. 
We provide the repository of our dataset and code used in the paper for reproducibility\footnote{\url{https://github.com/srijankr/casper}}.

	\section{Related Work}
	\label{sec:related_work}
	
\begin{table}[t]
    \footnotesize
	\centering
	\caption{Comparison of our proposed method (\method) against existing methods to measure perturbation and model stability.
	}
	\begin{tabular}{c|c|ccccccc}
		\toprule
		& \rotatebox[origin=l]{90}{\parbox{1.4cm}{\textbf{\method \\ (Proposed)}}} 
        &   \rotatebox[origin=l]{90}{\parbox{1.4cm}{Rev.Adv.~\cite{Revisiting_RecSys}}}  
		&   \rotatebox[origin=l]{90}{\parbox{1.4cm}{LOKI~\cite{zhang2020practical}}}  
		&  \rotatebox[origin=l]{90}{\parbox{1.4cm}{S-attack~\cite{fang2020influence}}} &  \rotatebox[origin=l]{90}{\parbox{1.4cm}{PoisonRec~\cite{PoisonRec}}} &  \rotatebox[origin=l]{90}{\parbox{1.4cm}{CF-attack~\cite{NIPS2016_6142} }}  &
		\rotatebox[origin=l]{90}{\parbox{1.4cm}{RL-attack~\cite{RL_attack}}} 
		& 
		\rotatebox[origin=l]{90}{\parbox{1.4cm}{LOO-user~\cite{black2021leave}}}
		 \\
		\midrule
				Deep Sequential Recommendation & \checkmark &  \checkmark & \checkmark &  &  \checkmark &  & \checkmark & \\
		Training Data Perturbation & \checkmark &  \checkmark &  \checkmark & \checkmark  &  \checkmark &  \checkmark & \checkmark & \checkmark\\
		Gray- or Black-box Perturbation & \checkmark & \checkmark   &  \checkmark   &  &   \checkmark  &  \\
		Interaction-level Perturbation & \checkmark &   &  &   &   & & &   \\
		Investigating Model Stability & \checkmark  &   &  &  &  &  & & \checkmark \\
		\bottomrule
	\end{tabular}
	\label{tab:comparators}
\end{table}

\noindent \textbf{Data Perturbation in Recommender Systems.}
Our work is broadly related to a stream of research on perturbations in deep recommender systems~\cite{Zhang2020PracticalDP, fang2020influence, PoisonRec, fang2018poisoning,taamr, oblivious_recsys, liu2019data, zhang2021data, wu2021triple, wu2021fight, yue2021black, anelli2021study, cohen2021black, liu2021adversarial}. Much of this work has generated perturbations that alter the rank of target item(s) (see Table~\ref{tab:comparators}). These methods highlight the vulnerability of specific recommendations. However, they provide incomplete stability since they focus on specific target items, rather than the entire or top-K rank lists for all users.\footnote{Setting all (top-K) items as targets can be inaccurate and computationally expensive.}
Furthermore, they are not appropriate as baselines because they are not applicable for interaction-level perturbations or work only on multimodal recommenders~\cite{anelli2021study, cohen2021black, liu2021adversarial} and matrix factorization-based models~\cite{wu2021fight, fang2020influence, wu2021triple}. 
Some CF-based~\cite{NIPS2016_6142} and RL-based~\cite{RL_attack} recommender systems, provide untargeted perturbations for recommender systems that reduce the model's prediction accuracy considerably. 
%\citet{black2021leave} study the impact of inserting or removing a single user (``leave-one-out'') on fairness of several ML models.
However, those methods do not work on our perturbation setting since they provide user- or item-level perturbations instead of interaction-level or focus on degrading the model's prediction accuracy without altering the rank lists of all users.

\noindent \textbf{Data Perturbation in Other Domains.}
Many input perturbation methods~\cite{FGSM, onepixel, Deepfool, guo2019simple, universal_attack, sarkar2017upset, zheng2016improving} have been developed for image classification. 
These methods cannot be directly applied to recommender systems due to complexities of sequential data (e.g., discrete input data and temporal dependency between interactions).
Many data perturbation algorithms~\cite{wallace2021concealed, chan-etal-2020-poison, kurita-etal-2020-weight, moradi2021evaluating,he2021petgen} for natural language processing (NLP) have been proposed. We cannot employ them directly for our setting since they either are targeted perturbations, have different perturbation levels (e.g., word or embedding modifications), or cannot model long sequential dependencies.

\noindent \textbf{Stability \& Multiplicity in Machine Learning.}
Our work is also related to a stream of work on stability and multiplicity in machine learning~\cite{marx2020predictive, d2020underspecification, coston2021characterizing, ali2021accounting, renard2021understanding, pawelczyk2020counterfactual, lee2022diversify, black2021leave, watsondaniels2022probabilistic,hsiang2022rashomon}. Recent work in this area has shown that datasets can admit multiple nearly-optimal solutions that exhibit considerable differences in other desirable characteristics (e.g., predictions on specific data points, behavior to model shifts, counterfactual explanations).
For instance, \citet{black2021leave} study data multiplicity caused by inserting or removing a single user (``leave-one-out'') on several ML models.
\citet{marx2020predictive} demonstrate the potential fairness issue in recidivism prediction problems.
While the majority of papers focus on the model multiplicity in classification models, they do not study the stability in recommender systems caused by data perturbations.

\section{Preliminaries}
	\label{sec:preliminary}
	We consider a sequential recommendation task, where a recommender model $\mathcal{M}: X \to R^{X}_{\mathcal{M}}$ is trained to learn users' behavioral patterns from a sequence of their actions. 
A trained model $\mathcal{M}$ generates a rank list of all items $R^{X_k}_\mathcal{M}$ that a user may interact with given a test interaction $X_k \in X_{\mathit{test}}$. Items are ordered in terms of the likelihood of user interaction, and the system shows the top-K items from the rank list $R^{X_k}_\mathcal{M}[1:K]$ to each user.
We denote the set of users and items as $U$ and $I$, respectively.
We study the sensitivity of four methods to train a sequential recommendation model:

\noindent $\bullet$ \textbf{\lstm~\cite{hochreiter1997long}:}
	    given a sequence of items, it predicts the next item via Long Short-Term Memory (LSTM).\\
\noindent $\bullet$ \textbf{\tisasrec~\cite{Tisasrec}:}
		a recent self-attention based model that predicts the next item using the relative time intervals and absolute positions among previous items.\\ 
\noindent $\bullet$ \textbf{\jodie~\cite{JODIE}:}
		a coupled RNN-based recommendation model which predicts the next item via RNNs to learn user and item embeddings.
		\\
\noindent $\bullet$
\textbf{\ltcross~\cite{Ltcross}:}
		a gated recurrent unit (GRU)~\cite{cho-etal-2014-learning} based model which uses contextual features, like time difference between interactions. This model is used in YouTube~\cite{Ltcross}.

\subsection{Datasets}
We use four recommendation datasets from diverse domains summarized in Table~\ref{tab:dataset}. 
In each dataset, we filter out users with fewer than 10 interactions.\\
\noindent $\bullet$ LastFM~\cite{LastFM, ren2019repeatnet, guo2019streaming, lei2020interactive, jagerman2019people} includes the music playing history of users represented as (user, music, timestamp). \\
\noindent $\bullet$ Foursquare~\cite{yuan2013time, ye2010location, yuan2014graph, yang2017bridging} is a point-of-interest dataset represented as (user, location, timestamp).\\
\noindent $\bullet$ Wikipedia~\cite{Wikipedia, dai2016deep, JODIE, li2020dynamic, pandey2021iacn, niverthi2022characterizing, kumar2015vews} contains the edit records of Wikipedia pages represented as (user, page, timestamp). \\
\noindent $\bullet$ Reddit~\cite{Reddit, dai2016deep, JODIE, li2020dynamic, pandey2021iacn,kumar2018community} includes the posting history of users on subreddits represented as (user, subreddit, timestamp).

\begin{table}[t!]
\footnotesize
	\centering
	\caption{Recommendation datasets used in Sections~\ref{sec:stability}, \ref{sec:proposed_method}, \ref{sec:experiment}.
	}
	\centering
	\begin{tabular}{ l  r r r r r}
		\toprule
		\textbf{Name} & \textbf{Users} & \textbf{Items} & \textbf{Interactions} & \textbf{Descriptions}\\
		\midrule
		LastFM & 980 & 1,000 &  1,293,103 & Music playing history  \\
		Foursquare & 2,106 & 5,597  & 192,602 & Point-of-Interest check-in\\
		Wikipedia & 1,914 & 1,000  & 142,143 & Wikipedia page edit history\\
		Reddit & 4,675 & 953 & 134,489 & Subreddit posting history\\
	\bottomrule
	\end{tabular}	
	\label{tab:dataset}
\end{table}

\subsection{Next-Item Prediction Metrics}
\label{sec:method:next_item_metrics}
The dataset-level performance of a sequential recommendation model is evaluated in a next-item prediction task by calculating the rank of the ground-truth item among all items, averaged over all test interactions. Two metrics are widely used: (i) Mean Reciprocal Rank (MRR)~\cite{voorhees1999trec}; (ii)  Recall@K (typically K=10)~\cite{JODIE,hansen2020contextual}. 
Both metrics lie between 0 and 1, and higher values are better. 
We refer to these two metrics as \textit{next-item metrics} as they provide average statistics of the ranks of ground-truth next items.

\section{Measuring Rank List Sensitivity}
\label{sec:method:ranklist_metrics}
We create a framework to measure the stability of recommendation systems against perturbations. 

\textbf{\textit{Procedure.}} First, we train a recommendation model $\mathcal{M}$ with the original data without perturbations, and it generates one ranked recommendation list $R^{X_k}_{\mathcal{M}}$ for each test interaction $X_k$ in test data $X_{\mathit{test}}$, where $k$ indicates an index of an interaction. 
Second, we train another recommendation model $\mathcal{M'}$ perturbed training data, and it generates ranked recommendation lists $R^{X_k}_\mathcal{M'}$, $\forall X_k \in X_{\mathit{test}}$.

We measure the similarity of recommendations for each test example $X_k$ by comparing the two recommendation lists $R^{X_k}_\mathcal{M}$ and $R^{X_k}_\mathcal{M'}$. We devise Rank List Sensitivity (RLS) metrics to measure the similarity (described in the next paragraph). 
Then, we average the individual RLS score across all $ X_k \in X_{\mathit{test}}$. 
If the model is perfectly stable, then $R^{X_k}_\mathcal{M}$ and $R^{X_k}_\mathcal{M'}$ should be identical $\forall X_k \in X_{\mathit{test}}$, and the average RLS value should be maximized. 

We repeat the above process multiple times with different random seeds (to average the impact of individual experiments), and report average values of RLS metrics across  different runs.
The average RLS value quantifies the model stability.

\textbf{\textit{Rank List Sensitivity Metrics.}}
To measure the stability of a recommendation model $\mathcal{M}$, we need metrics that can compare the similarity between recommendation lists generated with versus without perturbations, i.e., $R^{X_k}_\mathcal{M'}$ versus $R^{X_k}_\mathcal{M}$. 
Standard next-item prediction metrics (described in Section~\ref{sec:method:next_item_metrics}) only measure the rank of the ground-truth next item in one recommendation list. 
Extensions of these metrics, e.g., the difference in MRR or Recall, to measure similarity are not appropriate since these metrics can remain unchanged if the ground-truth item's rank is the same in $R^{X_k}_\mathcal{M}$ and $R^{X_k}_\mathcal{M'}$, even though the positions of the other items in the two rank lists are drastically different.
This happens in practice -- see Figure~\ref{tab:next_item_metrics}, where the difference between MRR and Recall values of recommendation models $\mathcal{M}$ and $\mathcal{M'}$ are almost identical. 
To compute the list similarity accurately, we need to measure how a perturbation impacts the order of \textit{all items} across two recommendation lists.
Thus, we need metrics that are sensitive to differences in the positions of all items, not only the ground-truth item. 

We introduce a formal metric called \textit{Rank List Sensitivity (RLS)} to quantify the stability of recommender systems by comparing the items and their ranking in two lists (or two top-K lists). 
Mathematically, RLS metrics of a model $\mathcal{M}$ against input perturbation are defined by the following:
$$RLS = \frac{1}{|X_{\mathit{test}}|}\sum_{\forall X_k \in X_{\mathit{test}}} \mathit{sim}(R^{X_k}_\mathcal{M}, R^{X_k}_\mathcal{M'})$$
where $\mathit{sim}(A,B)$ is a similarity function between two rank lists $A$ and $B$. We use the following two similarity functions in this paper.

\noindent (1) {\textbf{RBO (Rank-biased Overlap):} RBO~\cite{RBO} measures the similarity of orderings between two rank lists $R^{X_k}_\mathcal{M'}$ and $R^{X_k}_\mathcal{M}$. RBO lies between 0 and 1. Higher RBO means the ordering of items in the two lists is similar. For reference, the RBO between two randomly-shuffled rank lists is approximately $0.5$. 
RBO is more responsive to similarities in the top part of two rank lists, meaning that it imposes higher weights on the top-K items. This property distinguishes RBO from other measures like Kendall's Tau~\cite{kendall1948rank}. 
RBO of two rank lists $A$ and $B$ with $|I|$ items is defined as follows.}
$$ RBO(A,B) = (1-p) \sum_{d=1}^{|I|}{p^{d-1}\frac{|A[1:d] \cap B[1:d]|}{d}}$$
where $p$ is a tunable parameter (recommended value: 0.9).

\noindent (2) \textbf{Top-K Jaccard similarity}: The Jaccard similarity~\cite{jaccard1912distribution} $$Jaccard(A,B) = \frac{|A \cap B|}{|A \cup B|}$$ is a normalized measure of similarity of the contents of two sets $A$ and $B$.  We use it to measure the similarity of items in the top-K recommendation lists generated with and without perturbations, i.e., $R^{X_k}_\mathcal{M'}[1:K]$ and $R^{X_k}_\mathcal{M}[1:K]$. The Jaccard score ranges from 0 to 1, and is 
agnostic to the ordering of items. A model is stable if its Jaccard score is close to 1. 
In all experiments, we set K = 10 to compare the top-10 recommendations ~\cite{JODIE, hansen2020contextual}.

Top-K Jaccard metric can be useful for the industry due to its fast computation compared to RBO; RBO can be used for detailed analyses of the model stability since it focuses on full ranked lists.

	\section{Stability against Random Perturbations}
    \label{sec:stability}
    In this section, we investigate the stability of recommendation models against random perturbations. 

\textbf{\textit{Interaction-level Perturbations.}}
We measure the stability of a model with respect to arbitrary errors and noise through \textit{minimal random perturbations}. These perturbations change one randomly-chosen sample in the training data -- i.e., an interaction of a single user rather than all interactions of a user or an item. In particular, an interaction is either deleted (leave-one-out), inserted, or the interaction's item is replaced with another random item.

\textbf{
\textit{Experimental Setup.}}
Our goal is to test the stability of diverse recommendation models against a random interaction perturbation.
We use the first 90\% of interactions of each user for training the recommendation model, and the rest are used for testing, which is a common
setting used in several papers~\cite{hu2019sets2sets, wang2019kgat, donkers2017sequential, meng2020exploring}.
For each model, we use the hyperparameters mentioned in their original publications.
Other hyperparameters are set as follows: the maximum training epoch is set to 50, a learning rate is set to 0.001, and the size of the embedding dimension is set to 128. 
For \lstm and \tisasrec, the maximum sequence length per user is set to 50.

\textbf{\textit{Procedure to Measure Stability.}} 
We follow the procedure described in Section~\ref{sec:method:ranklist_metrics} and use the two RLS Metrics to measure the stability of recommendation models against random perturbations.

\begin{figure}[t]
    \centering
    \begin{subfigure}{0.48\textwidth}
    \centering
    \includegraphics[width=8.5cm]{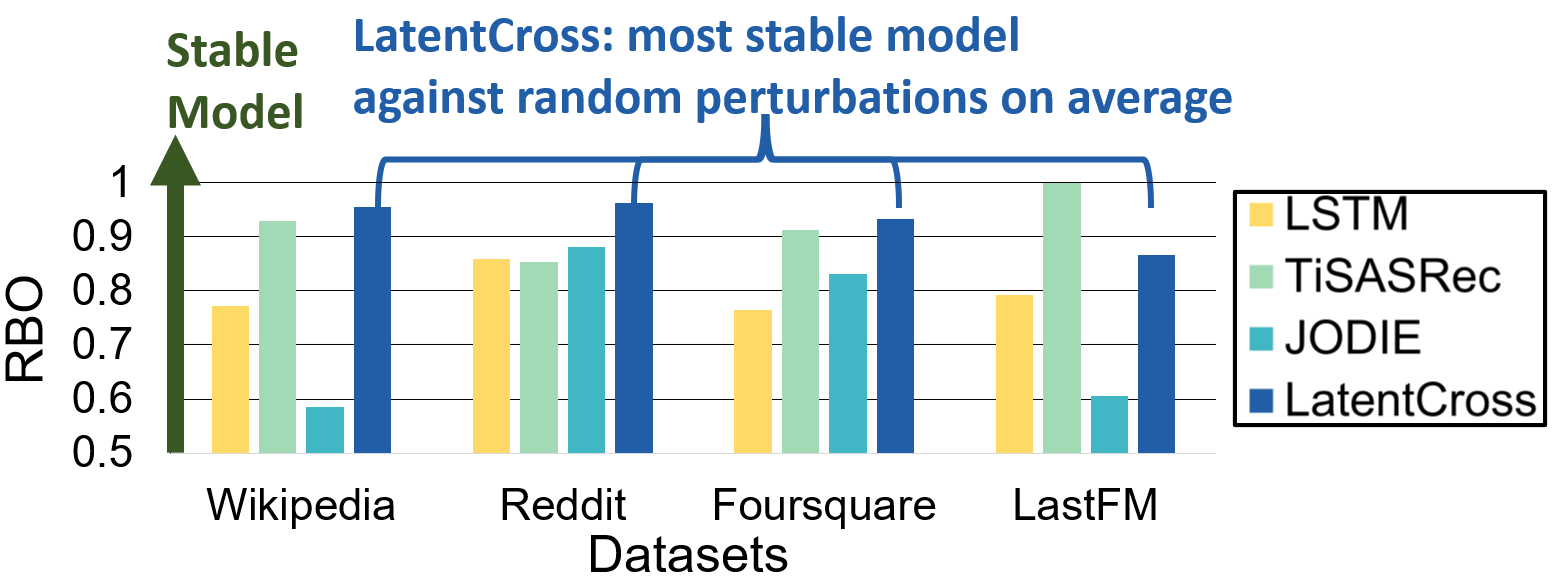}
    \caption{Random leave-one-out (LOO) perturbation}
    \label{fig:random_deletion}
    \end{subfigure}
    \begin{subfigure}{0.48\textwidth}
                \centering
         \includegraphics[width=8.5cm]{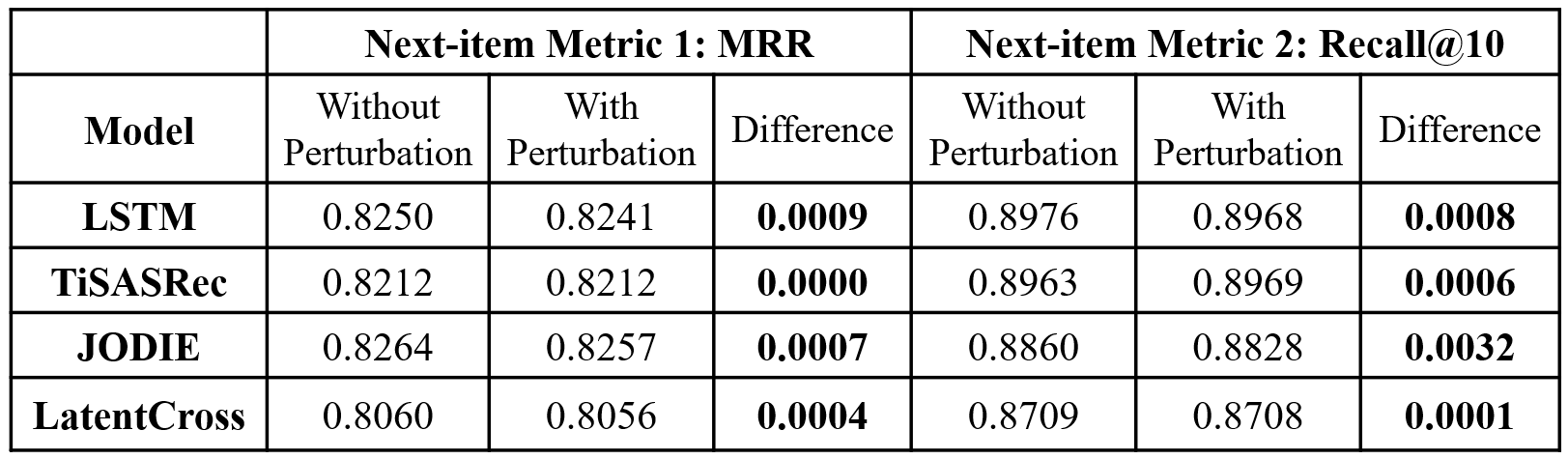}
            \caption{Next-item metrics of models against random LOO perturbation}
        \label{tab:next_item_metrics}
        \end{subfigure}
    \caption{(a) \textit{Stability of four recommendation models against random LOO perturbation}. Existing models exhibit unstable predictions since RBO scores after the perturbation are low.
    (b) \textit{Impact of random LOO perturbation on next-item predictions of recommendation models.} The differences in metrics with and without perturbations are marginal. 
    }
    \label{fig:random_perturbation}
\end{figure}

\textbf{\textit{Findings.}}
We present the RBO scores of four recommendation models 
on Foursquare  against random leave-one-out perturbation in Figure~\ref{fig:random_deletion}.
We observe that all four recommendation models exhibit low RBO scores on all datasets, ranging from 0.75 to 0.95 in most cases, while sometimes dropping below 0.6. Recall that since the RBO score between two randomly-shuffled rank lists is approximately 0.5, it shows that the drop of RBO caused by perturbations is meaningful, but the rank list does not change randomly, which is expected. 
Similar drops are observed for top-10 Jaccard similarity and in the case of insertion and replacement perturbations. Insertion and replacement perturbation results are excluded due to space limitation.
Thus, we observe the instability of existing models against even minor random perturbation.
Notably, perturbation of a user's interaction leads to drastic changes in the recommendations of unrelated users. 

Comparing the four models, \ltcross has the highest RBO in most cases against random perturbations.
This indicates that \ltcross is the most stable model against random perturbations.

\textbf{\textit{Controlling for Training Randomness while Measuring Model Stability.}}
Other research has found that randomness during the training (e.g., random initialization, mini-batch shuffling, etc.) can generate different models and predictions in machine learning~\cite{marx2020predictive, ali2021accounting}. 
Thus, in all our experiments (including the ones above), we specifically test the effect of the input data perturbation on model stability by controlling all other randomness (e.g., fixing the random seed and initialization). During a single run, we train two recommendation models $\mathcal{M}$ and $\mathcal{M'}$ (before and after perturbations) using \textit{the exact same settings without any training randomness}.
In other words, if there is no perturbation, the trained models $\mathcal{M}$ and $\mathcal{M'}$ and their outputs will be identical in every way. 

\textbf{\textit{Impact of Perturbations on Next-Item Prediction Metrics:}}
We find that the trained recommender models with and without perturbations have similar dataset-level performance metrics (both have almost identical MRR and Recall scores), as shown in Table~\ref{tab:next_item_metrics}. 
However, the generated recommendation lists are drastically different, as indicated by the low RBO and Jaccard scores.
This shows that multiple equivalent models can be trained that have similar dataset-level metrics, but provide conflicting recommendations. Similar findings have been made in other prediction settings~\cite{marx2020predictive}.

One may wonder that if the dataset-level metrics are the same, is there any concern if the rank lists vary? We argue that this is indeed a matter of concern due to the following three reasons: \\
(a) Since several equivalent models generate different predictions, the specific recommendations, e.g., which drug to administer or which treatment procedure to follow, can vary depending on which model is used. 
It is important for the algorithm designer and the end-user to know that if the recommendation for a certain user can be easily changed by unrelated minor perturbations, then perhaps none of the recommendations should be followed for that user. \\
(b) Since multiple recommender models exist with equivalent next-item prediction performance, then how can the algorithm designer decide which model to deploy? We argue that given comparable models, stabler recommender models should be used. \\
(c) Our work highlights the importance of ``beyond-accuracy'' metrics (e.g., RLS metrics) given that different recommender models vary in their stability with respect to the RLS metrics.

\textbf{\textit{Why are models unstable against minimal random perturbations?}} 
Only one interaction over one million interactions (size of the datasets used) is perturbed. Yet, it changes the rank lists and top-10 recommendation lists of all users. Why is there such a profound effect? This is due to two reasons.\\
\noindent (1) The slight change in training data leads to changes in the parameters of a trained recommendation model $\mathcal{M}$. 
Say an interaction in a mini-batch $m$ was perturbed. 
When processing $m$, model parameters $\Theta(\mathcal{M})$ will be updated differently during training (compared to when there is no perturbation). 
The changes in $\Theta(\mathcal{M})$ will affect the updates in later mini-batches. 
The differences will further cascade and multiply over multiple epochs. 
Thus, with perturbations, the final $\Theta(\mathcal{M})$ will be different from the ones obtained without perturbations, which can result in different rank lists. \\
\noindent (2) The model $\mathcal{M}$ is trained to accurately predict only the ground-truth next item as high in the rank list as possible (ideally, rank 1). 
However, $\mathcal{M}$ is not trained to optimize the positions of the other items in the rank list. 
Thus, the ordering of all except the ground-truth next item is highly likely to change due to input perturbation.

	\section{Stability against \method perturbation}
	\label{sec:proposed_method}
	While random perturbations show the model instability introduced due to arbitrary errors and noise, it is essential to find perturbations that can lead to even higher instability, which helps understand the lowest stability exhibited by a model. 
Adversaries can potentially exploit such perturbations to conduct untargeted attacks and make the recommendations unstable for all users. 
Thus, in this section, we ask: \textbf{\textit{which interaction should be perturbed to yield maximum instability in a recommendation model?}}
We aim to find perturbations that maximally change the rank lists $R^{X_k}_{\mathcal{M}}$ compared to $R^{X_k}_{\mathcal{M'}}$ $\forall X_k \in X_{\mathit{test}}$. 
As before, we will consider \textit{minimal interaction-level perturbations}, allowing one interaction to be perturbed. \textit{Three types of perturbations} can be made: leave-one-out (LOO), insertion, and replacement.%
Due to space constraints, we will highlight LOO perturbation results as other perturbations yield similar model instability.
Finally, we will consider \textit{gray-box perturbations} --- we assume access to training data and some model information such as the maximum sequence length of past user actions that the recommendation model uses to make predictions. Note that we do \underline{\textit{not}} require any details of the recommendation model such as the model's architecture, parameters, or  gradients~\cite{yue2021black,fang2018poisoning, huang2021data}.

\begin{figure}[t!]
    \centering
     \begin{subfigure}{0.48\textwidth}
\centering
 \includegraphics[width=7.5cm]{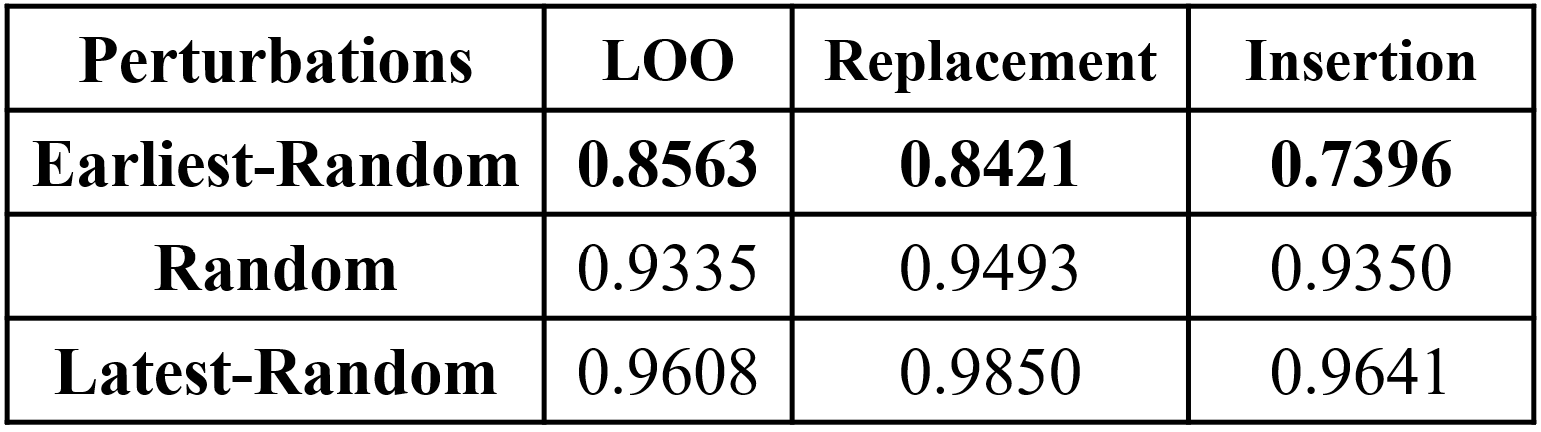}
\caption{Impact of perturbing interactions from different timestamps.}
\label{fig:hypothesis}
     \end{subfigure}
     \\
     \begin{subfigure}{0.48\textwidth}
     \centering
    \includegraphics[width=7.5cm]{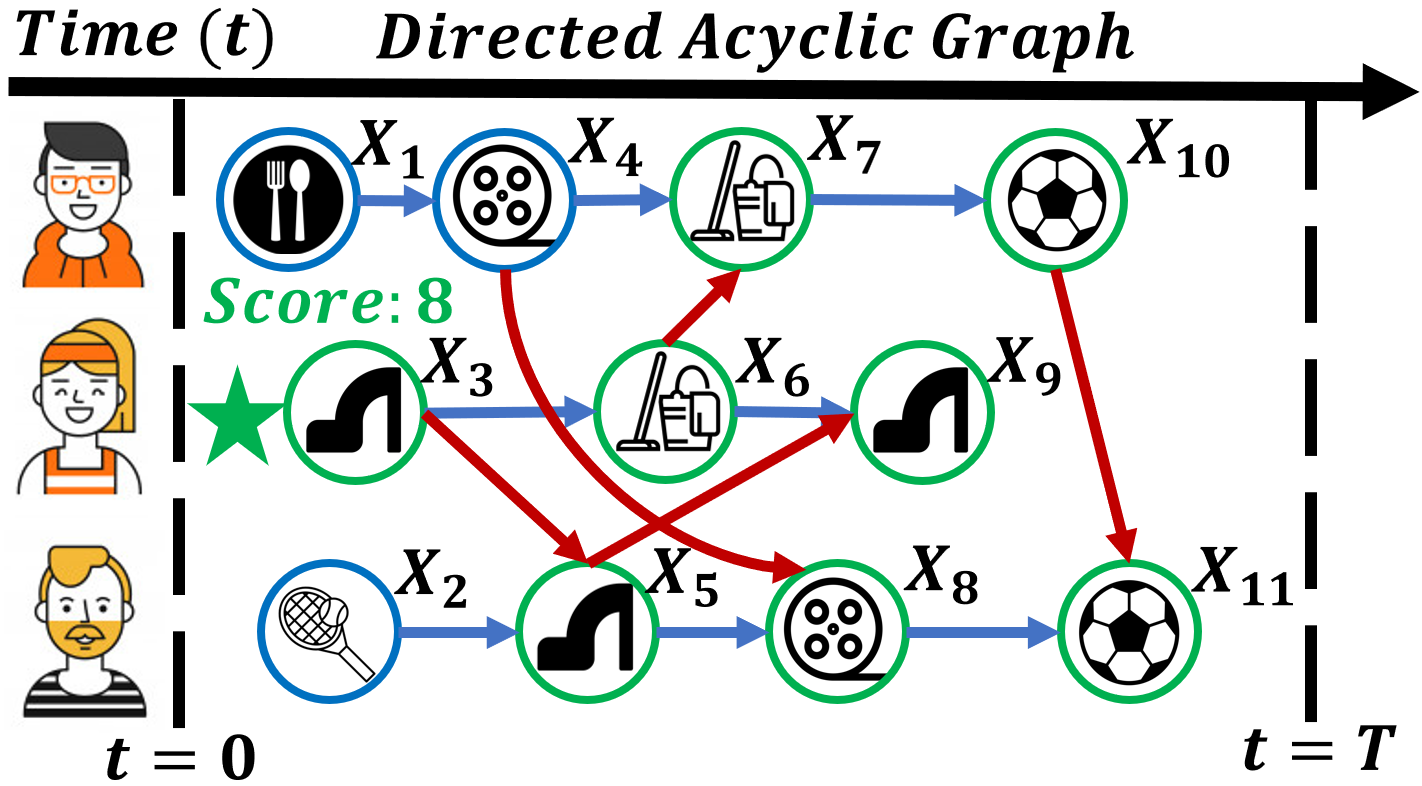}
    \caption{An IDAG corresponding to Figure~\ref{fig:main_figure}.}
	\label{fig:attack_plot}
	\end{subfigure}
	\caption{
	(a) Average RBO scores of perturbing interactions from different positions in the training data. Earliest-Random perturbation produces lower RBO than Random and Latest-Random perturbations. 
	(b) An IDAG corresponding to the interaction data in Figure~\ref{fig:main_figure}. 
    Blue and red edges indicate user- and item-sharing adjacent interactions, respectively. 
    Green-colored nodes (interactions) show all descendants (including itself) of an interaction $X_3$. 
    The cascading score of $X_3 = 8$, which is its number of descendants. 
    }
\end{figure}

\begin{algorithm*} [t!]
	\caption{\method: interaction-level perturbation based on cascading effect} \label{alg:main}
	\SetKwInOut{Input}{Input}
	\SetKwInOut{Output}{Output}
	\Input{ 
		Training interaction data $X_{\mathit{train}}$, users and items $U$ and $I$, training interaction sequences $X^{u}$ and $X^{i}$ (sorted by timestamp) for each user $u$ and item $i$\\
	}
	\Output{
    Perturbed training data $X_{\mathit{perturbed}}$\\
	}
	Initialize an interaction-to-interaction directed acyclic graph (IDAG) $G$ with all training interactions $X_{k} \in X_{\mathit{train}}$ as nodes \\
	\For(\Comment*[f]{\textbf{Creating edges in the IDAG $G$}}){each user $u \in U$}{ 
	    \For(\Comment*[f]{\textbf{Adding edges between consecutive interactions of $u$}}) {$k \in [1, 2, \ldots, |X^{u}|-1]$}{ 
	       Create an edge \textbf{from $X^{u}_{k}$ to $X^{u}_{k+1}$} in $G$\\
	    }
	}
	
	\For(\Comment*[f]{\textbf{Creating edges in the IDAG $G$}}){each item $i \in I$}{ 
	    \For(\Comment*[f]{\textbf{Adding edges between consecutive interactions of $i$}}) {$k \in [1, 2, \ldots, |X^{i}|-1]$}{ 
	        Create an edge \textbf{from $X^{i}_{k}$ to $X^{i}_{k+1}$} in $G$\\
	    }
	}
	{
	\For(\Comment*[f]{\textbf{Compute cascading scores $score(X_k)$, $\forall X_k \in X_{\mathit{train}}$}}){each interaction $X_{k} = (u,i,t) \in X_{\mathit{train}}$}{
	    \If{$Indegree(X_{k})==0$}{
	        Perform breadth-first search (BFS) starting from $X_{k}$ to find all the descendants of $X_k$ in $G$\\
	        $score(X_{k}) \longleftarrow$ total number of descendants of $X_{k}$ in $G$ \\
	    }
	}
	}
	\Comment{\textbf{Perturb the interaction with the highest cascading score}}{
    To obtain new perturbed training data $X_{\mathit{perturbed}}$, perturb the interaction $X_{\mathit{opt}} = \argmax{\forall X_{k} \in X_{\mathit{train}}}(score(X_{k}))$ \\
    }
\end{algorithm*}

\subsection{Perturbing Interactions from Different Timestamps}
A brute-force technique that tests the impact of every interaction perturbation on model stability is computationally prohibitive due to the need to retrain the model after each perturbation. 
To find an effective perturbation in a scalable manner, we first investigate the impact of perturbations in different positions in the training data. 

We take inspiration from an idea from temporal recommendation models~\cite{Ltcross, JODIE, hochreiter1997long}, where mini-batches $\mathcal{B} = \{B_1, \ldots, B_{T}\}, B_1 \cup \cdots \cup B_{T} = X_{\mathit{train}}$ are created in temporal order (i.e., first $P$ interactions in the first batch $B_1$, and so on). 
In such models, earlier batches contain training interactions with early timestamps, and perturbing an interaction in the earlier batches is equivalent to perturbing an interaction with early timestamps. 
Since we saw in the case of random perturbations that the impact of perturbations on model parameters can cascade, we ask: \textbf{\textit{how does perturbing interactions from different timestamps impact model stability?}}

We devise and compare three following heuristic perturbations: an \textit{Earliest-Random perturbation}, where the first interaction of a randomly selected user is perturbed, 
a \textit{Latest-Random perturbation}, where the last interaction of a randomly selected user is perturbed,
and a \textit{Random perturbation}, where a random training interaction is perturbed. 
We test this cascading effect on \ltcross model since it was the most stable against random perturbation. 

We use the RBO metric to measure RLS caused by these perturbations on the \ltcross model and Foursquare dataset (the hardest-to-predict dataset as per next-item metrics).
We perform each perturbation 10 times (randomly perturbing one interaction only each time). The resulting RBO score distributions are compared using
the Wilcoxon signed-rank test~\cite{wilcoxon1992individual}.

The RBO scores are shown in Table~\ref{fig:hypothesis}. 
Earliest-Random perturbation leads to the lowest RBO score in all three types of perturbations, i.e., LOO, replacement, and insertion (all p-values \textless 0.05). 
We also observe that between Random and Latest-Random, the former has lower RBOs.
These findings show that perturbing earlier timestamp interactions leads to higher instability in recommendations.
Since this happens due to the cascading impact of model parameter changes over mini-batch updates, we call this a ``\textit{cascading effect}''.

\subsection{\method: Interaction-level Perturbation based on Cascading Effect}

Now, we leverage the cascading effect to propose a new perturbation, named \method (\underline{Cas}cade-based \underline{Per}turbation). 

To approximate the impact of perturbing an interaction $X_k$, we define a \textit{cascading score} of $X_k$ as the number of training interactions that will be affected if $X_k$ is perturbed. 
Inspired by temporal recommendation models~\cite{Ltcross, JODIE, hochreiter1997long}, we create an interaction-to-interaction dependency graph, which encodes the influence of one interaction on another. 
Then, we approximate the cascading score of interaction $X_k$ as the number of descendants of $X_k$ in this graph. 
\method aims to identify the training interaction which has the highest cascading score, since its perturbation would maximize the cascading effect. 
Algorithm~\ref{alg:main} shows the key steps of the method. 

\noindent
\textbf{\textit{Creating the interaction-to-interaction dependency DAG:}} 
We create a graph-based technique to approximate an interaction's cascading score without retraining the recommendation model. 
We first construct an interaction-to-interaction dependency directed acyclic graph (IDAG; lines 1-7 in Algorithm~\ref{alg:main}), where nodes are training interactions and directed edges represent which interaction influences another. 
The edge encodes the dependency that if the $k^{th}$ interaction of user $u$ (or item $i$) is perturbed, it will influence the $k+1^{th}$ interaction of user $u$ (or item $i$). 
The IDAG corresponding to the training interactions from Figure~\ref{fig:main_figure} is presented in Figure~\ref{fig:attack_plot}.
Two nodes in the IDAG are connected by a directed edge if they are either consecutive interactions of the same user (e.g., $X_1$ and $X_4$) or of the same item (e.g., $X_3$ and $X_{5}$). 
A directed edge must follow the temporal order from early to later timestamp. 
No edges are present between nodes with the same timestamp. 
Thus, each node has at most two outgoing edges (first to the next interaction of the user and second to the next interaction of the item).
If the recommendation model has a maximum sequence length ($L$), the IDAG is constructed only with the latest $L$ interactions of each user.

\noindent
\textbf{\textit{Calculating the cascading score in IDAG:}} The cascading score of a node $X_k$ is approximated as the total number of descendants of $X_k$ in the IDAG. 
Descendants of a node $X_k$ in the IDAG are defined as all the nodes reachable from $X_k$ by following the outgoing edges in the IDAG. 
For example, in Figure~\ref{fig:attack_plot},   $X_3$ has 8 descendants (including itself), the highest among all nodes. 
By definition, a node's parent will have a higher cascading score than the node itself. 
Hence, we accelerate the computation by calculating the cascading scores of zero in-degree nodes only (lines 8-11 in Algorithm~\ref{alg:main}).
Finally, \method perturbs the node with the highest cascading score since it would maximize the cascading effect (line 12 in Algorithm~\ref{alg:main}).

We have theoretically and experimentally shown that \method scales near-linearly to the dataset size (Section~\ref{sec:method:complexity} and Figure~\ref{fig:runtime_analysis}). 

\subsection{Complexity Analyses of \method}
\label{sec:method:complexity}
We analyze the time and space complexities of \method. 
We assume the maximum sequence length of a model is $L$.

\noindent
{\textbf{Time complexity.}} \method first trains and tests a given recommendation model $\Theta$ with original input data, which takes $\mathcal{O}(\mathcal{T}(\Theta))$, where $\mathcal{T}(\Theta)$ is the time complexity of $\Theta$. After that, \method constructs the IDAG which takes $\mathcal{O}(|U|L)$ where $|U|$ is the number of users. Computing cascading scores of zero in-degree nodes in the IDAG, which takes $\mathcal{O}(Z|U|L)$ where $Z$ is the number of zero in-degree nodes in the IDAG. Perturbing an interaction with the highest cascading scores takes $\mathcal{O}(Z)$.
Finally, \method retrains the model $\Theta$ with perturbed data and computes RLS metrics, which takes  $\mathcal{O}(\mathcal{T}(\Theta)+N_{\mathit{test}}|I|)$ since RBO should be calculated with all items $|I|$, where $N_{\mathit{test}}$ is the number of test interactions. The final time complexity of \method is  $\mathcal{O}(\mathcal{T}(\Theta)+N_{\mathit{test}}|I|+Z|U|L)$.

\noindent
{\textbf{Space complexity.}} The first step of \method is training and testing a deep sequential recommendation model $\Theta$ with original input data, which takes $\mathcal{O}(\mathcal{S}(\Theta)+N_{\mathit{test}}|I|)$ space since we need to store original rank lists for all test interactions, where $\mathcal{S}(\Theta)$ is the space complexity of $\Theta$. After that, \method constructs the IDAG which takes  $\mathcal{O}(|U|L)$ space. The next step is computing cascading scores of zero in-degree nodes in the IDAG, which takes $\mathcal{O}(|U|L)$ space. Finally, \method retrains the model $\Theta$ with perturbed data and computes RLS metrics, which takes  $\mathcal{O}(\mathcal{S}(\Theta)+N_{\mathit{test}}|I|)$ space. The final space complexity of \method is $\mathcal{O}(\mathcal{S}(\Theta)+N_{\mathit{test}}|I|+|U|L)$.
	
	\section{Experimental Evaluation of \method}
	\label{sec:experiment}
	In this section, we evaluate \method by the following aspects. 

\noindent (1) \textbf{Stability of Recommendation Models against Diverse Perturbations (Section~\ref{sec:exp:best_attack}).}
	How stable are existing recommender systems against \method and baseline perturbations?

\noindent (2) \textbf{Impact of Perturbations on Different Users (Section~\ref{sec:exp:fairness}).}
Are there any user groups that are more susceptible and sensitive to input data perturbations?
	
\noindent (3) \textbf{Impact of the Number of Perturbations (Section~\ref{sec:exp:num_perturbation}).}
	Is the performance of \method  proportional to the number of  perturbations allowed on the dataset?

\noindent (4) \textbf{Running Time Analysis (Section~\ref{sec:exp:runtime}).}
	Does the running time of \method scale with the dataset size?
	
\begin{table*}[t!]
\caption{\textit{Effectiveness of perturbations on Foursquare dataset.} We find instability of existing recommendation models measured by the RBO metric against LOO (left) and item replacement perturbations (right). 
All RBO scores are lower than 1.0. 
The {\color{blue!75}best perturbation} in each column is colored {\color{blue!75}blue}, and the {\color{blue!35}second best is light blue}, in terms of achieving the lowest RBO score.
% \method is the best across almost all but one setting, in which it is the second best. 
}
\hspace{-5mm}
\begin{subtable}{0.49\textwidth}
\centering
\footnotesize
\caption{LOO perturbation comparison using RBO}
\begin{tabular}{|c|c|c|c|c|}
\hline
\textbf{\begin{tabular}[c]{@{}c@{}}Model / \\ Perturbations \end{tabular}}  & \textbf{\lstm} & \textbf{\tisasrec} & \textbf{\jodie} & \textbf{\ltcross} \\ \hline
\textbf{Random}                                                     &  0.7799        & 0.9117            & 0.8316         & 0.9335           \\ \hline
\textbf{\begin{tabular}[c]{@{}c@{}}Earliest-\\Random\end{tabular}} & 0.7876        & 0.8776        & 0.8211         & 0.8563           \\ \hline
\textbf{\begin{tabular}[c]{@{}c@{}}Latest-\\Random\end{tabular}} & 0.7763        & \cellcolor{blue!10} 0.8515        & 0.8420         & 0.9608           \\ \hline
\textbf{TracIn~\cite{pruthi2020estimating}}                                                     & \cellcolor{blue!10} 0.7733        &  0.8545            &  0.8713         & 0.9625           \\ \hline
\textbf{\begin{tabular}[c]{@{}c@{}}Rev.Adv.~\cite{Revisiting_RecSys} (random) \end{tabular}}                                                    & 0.7798        &      0.8955        & 0.8491       & 0.9317         \\ \hline
\textbf{\begin{tabular}[c]{@{}c@{}}Rev.Adv.~\cite{Revisiting_RecSys}  (earliest) \end{tabular}}                                                    & 0.7787        &      0.8911        & \cellcolor{blue!25} \textbf{0.7185}       & \cellcolor{blue!10} 0.7403        \\ \hline
\multicolumn{5}{|c|}{\textbf{Proposed method}}\\\hline
\textbf{\method}                                                     & \cellcolor{blue!25} \textbf{0.7709}     & \cellcolor{blue!25} \textbf{0.8450}         &   \cellcolor{blue!10} 0.7896             & \cellcolor{blue!25} \textbf{0.6662}           \\ \hline
\end{tabular}
\captionsetup{font={small}}
\label{tab:deletion_performance}
\end{subtable}
\hspace{5mm}
\begin{subtable}{0.49\textwidth}
\centering
\footnotesize
\caption{Item replacement perturbation comparison using RBO}
\begin{tabular}{|c|c|c|c|c|}
\hline
\textbf{\begin{tabular}[c]{@{}c@{}}Model  / \\ Perturbations \end{tabular}}  & \textbf{\lstm} & \textbf{\tisasrec} & \textbf{\jodie} & \textbf{\ltcross} \\ \hline
\textbf{Random}                                                     & 0.7795        & 0.9143            & 0.9414         &  0.9493           \\ \hline
\textbf{\begin{tabular}[c]{@{}c@{}}Earliest-Random\end{tabular}} & 0.7743        & 0.8886            & 0.8871         & 0.8421           \\ \hline
\textbf{\begin{tabular}[c]{@{}c@{}}Latest-Random\end{tabular}} & 0.7814        & 0.8553        & 0.8989         & 0.9850           \\ \hline
\textbf{TracIn~\cite{pruthi2020estimating}}                                                 & 0.7934        & 0.8520            & 0.9280         & 0.9696          \\ \hline
\textbf{\begin{tabular}[c]{@{}c@{}}Rev.Adv.~\cite{Revisiting_RecSys} (random) \end{tabular}}                                                    & 0.7856        &     0.8782        & 0.9159       & 0.9538         \\ \hline
\textbf{\begin{tabular}[c]{@{}c@{}}Rev.Adv.~\cite{Revisiting_RecSys}  (earliest) \end{tabular}}                                                    & 0.7747        &      0.9375       & 0.8257     & 0.7449        \\ \hline
\multicolumn{5}{|c|}{\textbf{Proposed method}}\\\hline
\textbf{\method (random)}                                                     &  0.7665    &  0.8482       &         0.6691        &   0.6065           \\ \hline
\textbf{\method (popular)}                                                     &  \cellcolor{blue!25} \textbf{0.7557}     &  \cellcolor{blue!10} 0.8477        &         \cellcolor{blue!10} 0.6114        &   \cellcolor{blue!10} 0.5435           \\ \hline
\textbf{\method  (unpopular)}                                                     &   \cellcolor{blue!10} 0.7615     &   \cellcolor{blue!25} \textbf{0.8471}        &        \cellcolor{blue!25} \textbf{0.5228}        &  \cellcolor{blue!25} \textbf{0.5193}         \\ \hline
\end{tabular}
\label{tab:replacement_performance}
\end{subtable}
\label{tab:reliability}
\end{table*}

\subsection{Experimental Settings}
\label{sec:exp:settings}

\subsubsection{Datasets}
\label{sec:datasets}
We use the four standard datasets introduced in Section~\ref{sec:stability}. LastFM is a widely used recommendation benchmark dataset~\cite{ren2019repeatnet, guo2019streaming, lei2020interactive, jagerman2019people}, Foursquare is broadly utilized for point-of-interest recommendations~\cite{yuan2013time, ye2010location, yuan2014graph, yang2017bridging}, and Wikipedia and Reddit are popular for social network recommendations~\cite{dai2016deep, JODIE, li2020dynamic, pandey2021iacn}. 
We select these datasets for experiments because (a) they come from diverse domains, thus ensuring generalizability, and (b) the timestamps of interactions reflect when the corresponding activities happened (as opposed to Amazon review datasets where a review is posted much after a product is purchased, or MovieLens review dataset where a review is posted much after a movie is watched).

\subsubsection{Baseline Methods}
\label{sec:competitors}
To the best of our knowledge, there are no interaction-level perturbation methods for existing recommendation models. Therefore, we create strong baselines and two state-of-the-art methods based on the broader literature as follows:

\noindent $\bullet$ \textbf{Random perturbation:}
		It randomly chooses an interaction for perturbation among all training interactions. 

\noindent $\bullet$ 
\textbf{Earliest-Random perturbation:}
		It randomly chooses an interaction for perturbation among the first interactions of all users in the training data. 
	
\noindent $\bullet$ \textbf{Latest-Random perturbation:} 
		It randomly chooses an interaction for perturbation among the last interactions of all users in the training data. 
	
\noindent $\bullet$ \textbf{TracIn~\cite{pruthi2020estimating} perturbation:} 
	It chooses the most important training interaction for perturbation, defined in terms of reducing the model's loss during training. 
		We use an influence estimator TracIn~\cite{pruthi2020estimating} that utilizes loss gradients from the model saved at every $T$ epoch to compute interaction importance. 
	
\noindent $\bullet$ \textbf{Rev.Adv.~\cite{Revisiting_RecSys} perturbation:} It inserts a fake user with interactions crafted via a bi-level optimization problem for perturbations. To adapt it for our leave-one-out (LOO) and replacement perturbation settings, we first find the most similar user in the training data to the fake user, and perform LOO or item replacement of the earliest or random interaction of that user, respectively. Therefore, we create two versions of Rev.Adv. --  Rev.Adv.~\cite{Revisiting_RecSys} (random) and Rev.Adv.~\cite{Revisiting_RecSys} (earliest), which indicates the method chooses a random or earliest interaction of a user for perturbation, respectively.
	
Note that we do not include baselines that work only on multimodal recommenders~\cite{anelli2021study, cohen2021black, liu2021adversarial} and matrix factorization-based models~\cite{wu2021fight, fang2020influence, wu2021triple} as these are not applicable to our setting. We have also not included baselines that have shown similar or worse performance~\cite{yue2021black, zhang2020practical, PoisonRec} compared to the above baselines, particularly compared to Rev.Adv.~\cite{Revisiting_RecSys}. 
In replacement and insertion perturbations, the new item can be selected using three different strategies: selecting an item randomly, selecting the most popular item, or selecting the least popular (i.e., unpopular) item.

\subsubsection{Recommendation Models}
\label{sec:exp:competitors}
We use popular recommender models: \lstm~\cite{hochreiter1997long}, \tisasrec~\cite{Tisasrec}, \jodie~\cite{JODIE}, and \ltcross~\cite{Ltcross} described earlier to test the effectiveness of \method and baselines.

\subsubsection{Experimental Setup}
\label{sec:exp:setup}
We follow the same experimental setup, as described previously in Section~\ref{sec:stability}. Additionally, we use the following settings. 
We repeat all experiments multiple times and report average values of RLS metrics.
To construct the IDAG for \method, we use all the interactions in \jodie and \ltcross. For \lstm and \tisasrec, we use the latest 50 interactions per user, as defined by the maximum sequence length in the original papers. 
To compute the influence of interactions in the TracIn perturbation, we take training loss gradients with respect to the last hidden layer. We save the loss gradients every 10 epochs and fix step sizes to the default learning rate of $0.001$.

\begin{figure}[t!]
    \vspace{4mm}
    \hspace*{-1cm}
    \begin{subfigure}{0.24\textwidth}
    \centering
    \includegraphics[width=4.5cm]{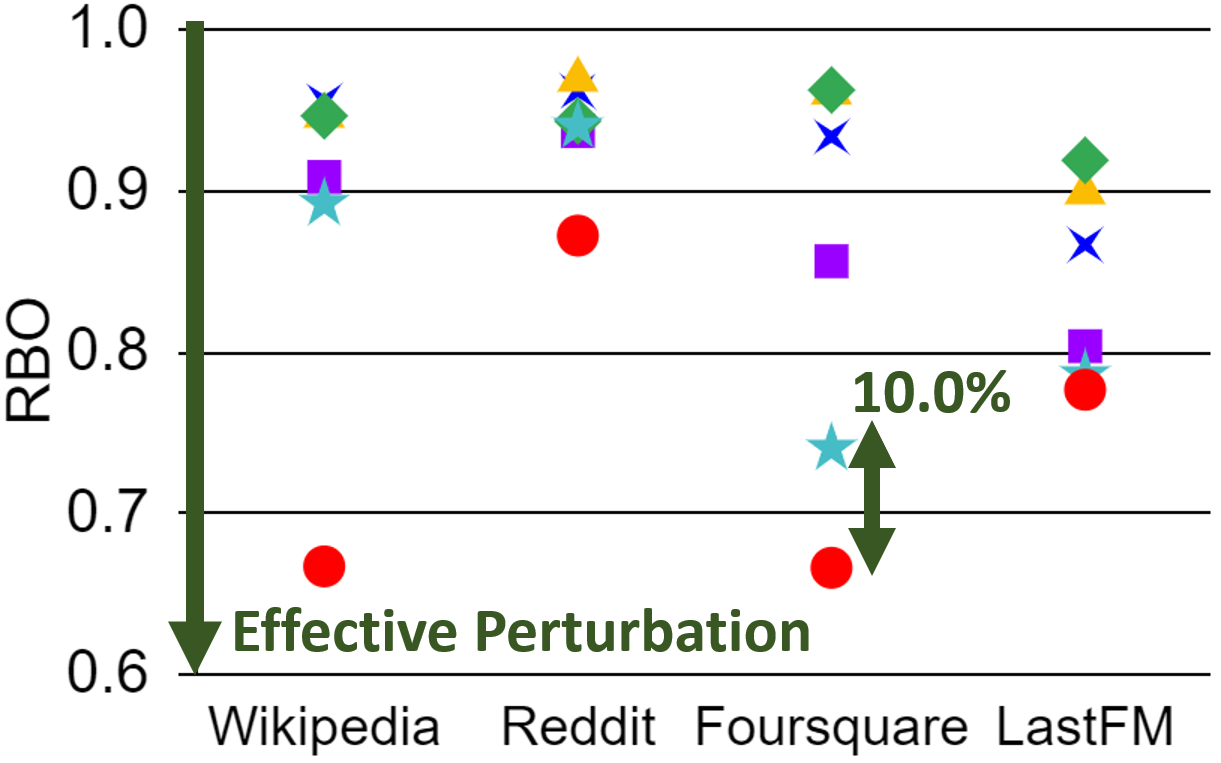}
    \captionsetup{justification=centering}
    \caption{RBO (LOO)}
    \label{fig:ltcross_deletion_RBO}
    \end{subfigure}
    \hspace*{2mm}
    \begin{subfigure}{0.24\textwidth}
    \centering
    \includegraphics[width=4.5cm]{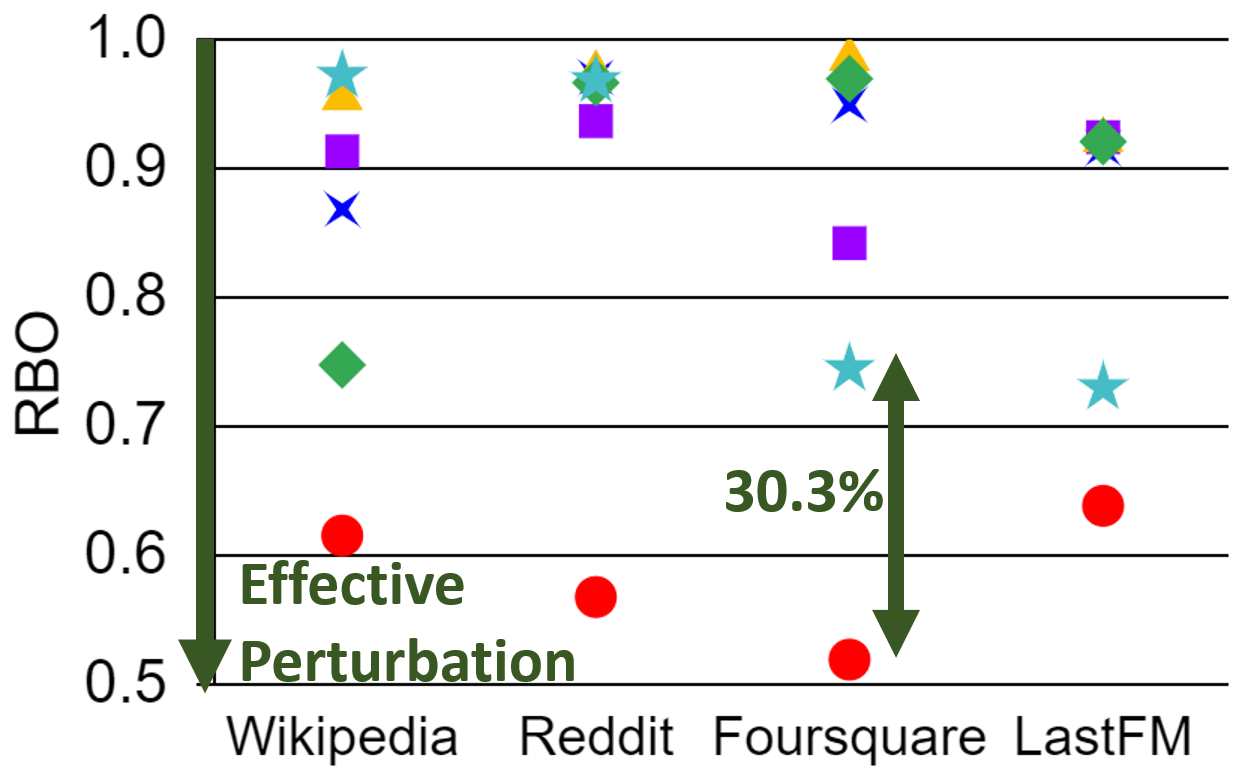}
    \captionsetup{justification=centering}
    \caption{RBO (item replacement)}
    \label{fig:ltcross_replace_RBO}
    \end{subfigure}
    \hspace*{-1cm}
    \begin{subfigure}{0.48\textwidth}
    \centering
    \includegraphics[width=9cm]{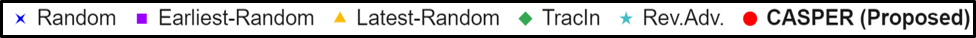}
    \end{subfigure}
    \caption{\textit{Comparing perturbations on \ltcross model across all datasets.}
    \method shows the best perturbation performance. 
    }
	\label{fig:ltcross_perturbations}
\end{figure}

\begin{figure}[t!]
    \vspace{4mm}
    \hspace*{-3mm}
    \begin{subfigure}{0.23\textwidth}
    \centering
    \includegraphics[width=4.2cm]{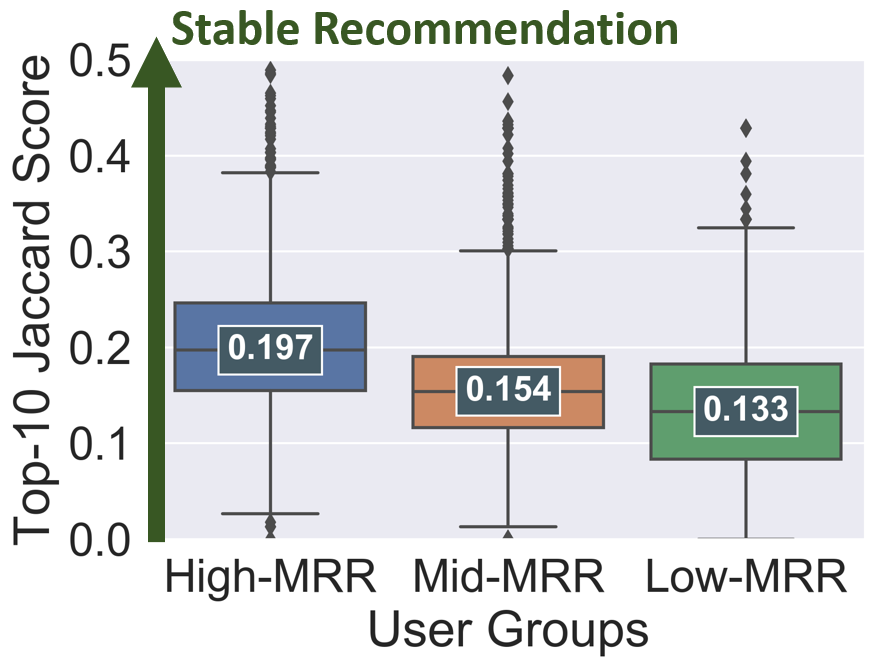}
    \captionsetup{justification=centering}
    \caption{Jaccard@10 on Foursquare}
    \label{fig:ltcross_foursquare_user_analysis}
    \end{subfigure}
    \hspace*{3mm}
    \begin{subfigure}{0.23\textwidth}
    \centering
    \includegraphics[width=4.2cm]{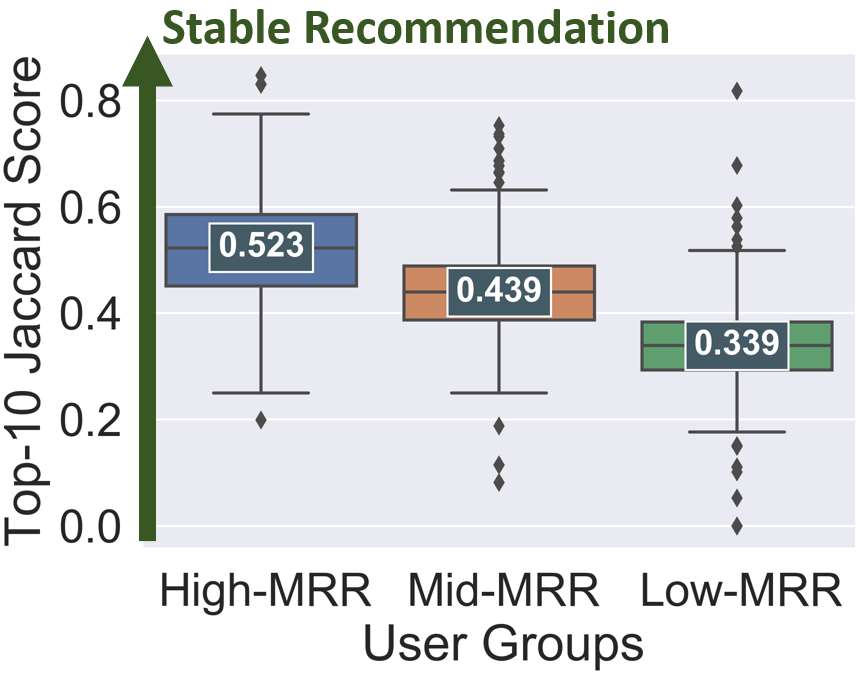}
    \captionsetup{justification=centering}
    \caption{Jaccard@10 on LastFM}
    \label{fig:ltcross_lastfm_user_analysis}
    \end{subfigure}
    \caption{\textit{Comparing impact of perturbations across user groups.}  Users with low accuracy receive more unstable predictions when \method perturbation is applied, which can cause a user fairness issue. This plot is for LOO perturbation results on \ltcross model. 
    }
	\label{fig:ltcross_analysis}
\end{figure}

\begin{figure*}[t!]
    \centering
    \includegraphics[width=16cm]{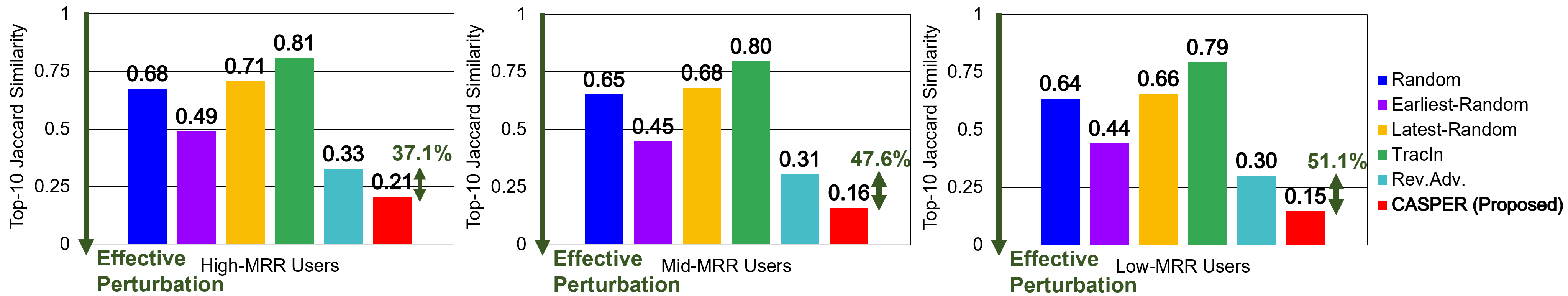}
\caption{\textit{Comparing different perturbation methods' impact across user groups}. We see LOO perturbation on \ltcross model and Foursquare dataset as per top-K Jaccard similarity (K=10), averaged over users with high-MRR (left), mid-MRR (middle), and low-MRR (right), respectively.
Users with low accuracy suffer more from training data perturbations with any perturbation method. 
\method leads to the highest reduction of top-10 Jaccard similarity and outperforms all baselines across all user groups. 
}
\label{fig:top10_jaccard}
\end{figure*}

\subsection{Stability of Recommendation Models against Diverse Perturbations}
\label{sec:exp:best_attack}
\textit{Perturbations of all models on Foursquare dataset.}
Table~\ref{tab:reliability} compares the performance of all perturbation methods on all four recommendation models and Foursquare dataset (the hardest-to-predict in terms of next-item metrics), averaged over 3 repetitions. Each column highlights the \textbf{{\color{blue!75}best}} and \textbf{{\color{blue!35}second-best}} perturbation model, in terms of the lowest RBO score.

We observe the instability of all recommendation models against LOO and replacement perturbations.
The RBO scores of all the recommendation models drop significantly below 1.0, indicating their low stability. 
\method achieves the best performance across all but one setting, where it performs the second best. It leads to the most reduction of RBO in most cases. \method shows lower variances of RLS values than those of baselines across different runs.
We observe that \method is more effective on \jodie and \ltcross models, since the other two models (\lstm and \tisasrec) use maximum sequence lengths, which limit their interactions' cascading effects.
In some cases, e.g., \jodie and \ltcross in item replacement, their resulting RBO drops close to 0.5, which is similar to the case of random shuffling of ranked lists. 
Similar observations hold with top-K Jaccard score and for insertion perturbations.
It is also worth mentioning that Rev.Adv. (earliest) outperforms Rev.Adv. (random) in most cases, which also substantiates the cascading effect.

In \textit{item replacement perturbation} (Table~\ref{tab:reliability}(b)), \method\ outperforms other methods in all cases. For \method, replacing the item with the least popular item is the most effective strategy among all the others.
One possible reason is that the change in user embeddings and model parameters by using an unpopular item will be the highest. 
Injection of the unpopular item diversifies the user's interactions and embedding the most, and model parameters can be updated most differently. 
This major update will cascade to later interactions and change all users' recommendations drastically.

\textit{Perturbations on \ltcross model on all datasets.}
We further evaluate the effectiveness of \method versus baselines on the \ltcross model (the most stable model against random perturbations) across  four datasets. 
The results are shown in Figures~\ref{fig:ltcross_deletion_RBO} and \ref{fig:ltcross_replace_RBO}. 
We confirm unstable predictions of \ltcross against \method LOO and item replacement perturbations as per RBO. Top-K Jaccard metric and insertion perturbations also show similar results. Notably, \method outperforms all baselines across all datasets on the \ltcross model.

Across all datasets, Latest-Random baseline performs worse than the Random, which performs worse than the Earliest-Random, due to cascading effect. Similarly, all random perturbations have worse performance than advanced perturbations like \method.

\subsection{Impact of Perturbations on Different Users}
\label{sec:exp:fairness}
To investigate the differential impact of training data perturbations on different users, we divide users into three groups: (1) High-MRR users, containing users who lie in the top 20\% according to average MRR, (2) Low-MRR users, containing users with the lowest 20\% average MRR, and (3) Mid-MRR users, which contains the remaining set of users. 
We contrast the average RLS of users across the three groups. 
Figures~\ref{fig:ltcross_foursquare_user_analysis} and \ref{fig:ltcross_lastfm_user_analysis} compare the top-10 Jaccard scores across the three user groups on \ltcross model and two datasets (Foursquare and LastFM) against \method LOO perturbation. 
We discover that the trend of stability follows the accuracy trend -- users with high accuracy receive relatively more stable predictions than the low-accuracy user group. 
This phenomenon highlights the relatively higher instability faced by users for which the model is already unable to make accurate predictions. 
This raises an aspect of unfairness across user groups. 
This highlights the need that practitioners should evaluate model stability across user groups before deploying models in practice.

Furthermore, we observe the same trend across different perturbation methods, as shown in Figure~\ref{fig:top10_jaccard}.
Regardless of the perturbation method, low-MRR users experience lower stability compared to the other two groups. 
Notably, \method is able to generate the lowest stability across all user groups. 
Addressing the differential impact across user groups will be  important to study in future work. 

\subsection{Impact of the Number of Perturbations}
\label{sec:exp:num_perturbation}
Intuitively, more perturbations in training data will cause higher instability of a model. 
To test the effect of the number of perturbations on \method, we increase the number of perturbations from 1 to 8 and check its LOO perturbation performance on \ltcross model and Foursquare dataset. \method selects $k$ interactions with the highest cascading score when the number of perturbations is $k$. As shown in Figure~\ref{fig:attack_scalability}, the performance of \method scales near-linearly with the number of perturbations. Replacement and insertion perturbations show similar trends.

\subsection{Running Time Analysis}
\label{sec:exp:runtime}
\begin{figure}[t!]
    \hspace{-3mm}
    \begin{subfigure}{0.23\textwidth}
    \centering
    \includegraphics[width=3.8cm]{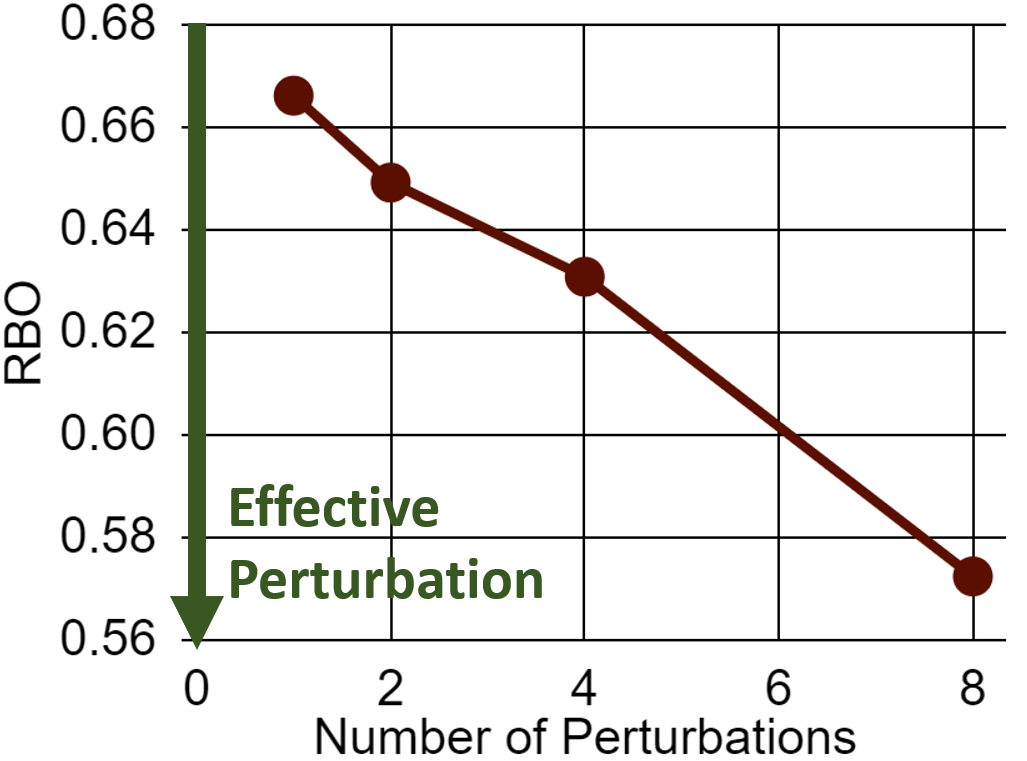}
     \phantomsubcaption
    \label{fig:attack_scalability}
    \end{subfigure}
    \hspace{2mm}
    \begin{subfigure}{0.23\textwidth}
    \centering
    \includegraphics[width=4cm]{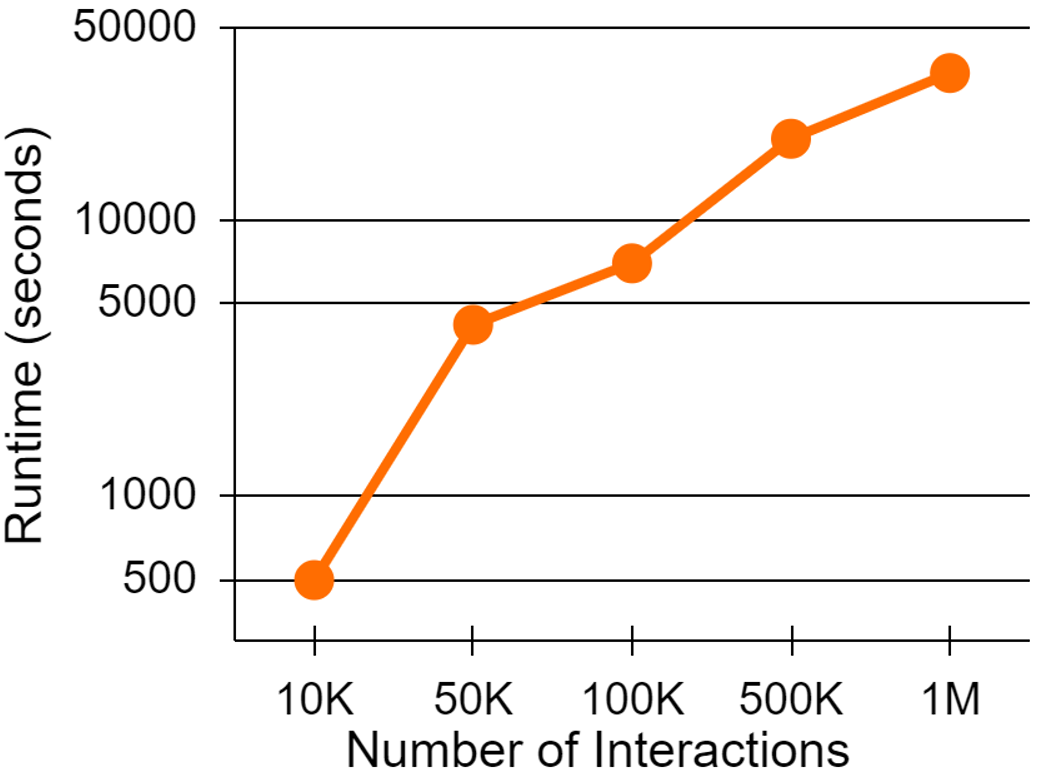}
     \phantomsubcaption
    \label{fig:runtime_analysis}
    \end{subfigure}
    \caption{\textit{(a) Perturbation scalability and (b) runtime} of \method. 
    }
	\label{fig:runtime_and_attack_scalability}
\end{figure}

We vary the number of interactions in a dataset to test whether the runtime of \method  is scalable to the input data size. Specifically, we measure the running time of LOO perturbation of \method on \ltcross model and LastFM dataset (the largest), while varying the number of interactions in the dataset from 10,000 to 1,000,000. Figure~\ref{fig:runtime_analysis} shows \method scales near-linearly with the dataset size. This empirically validates the time complexity of \method (see Section~\ref{sec:method:complexity}), which is linear as per the total number of interactions.

\vspace{-2mm}
	\section{Concluding Remarks}
	\label{sec:conclusion}
	Our work highlights that recommendation models can exhibit instability to minor changes in their training data. These effects underscore the need to measure this instability, and to develop methods that are robust to such changes. The measures and methods developed in this paper are an initial step in this direction. In particular, \method depends on cascading effect which is inspired by temporal recommendation models, meaning that it may return solutions that are sub-optimal for methods that are not trained with temporally-ordered mini-batches. 

Future work topics include: expanding \method to handle more complex perturbations, or to find more effective perturbations (e.g., interaction reordering) for other training regimes; and improving scalability of \method to handle very large interaction graphs (e.g., by creating approximations of cascading scores using a randomly-sampled interaction graphs, rather than the entire graph); developing methods that induce stability to data perturbations (e.g., via multi-objective learning aiming to accurately predict next items and preserve rank lists of a recommendation model simultaneously).
	
\vspace{-2mm}
	\section*{Acknowledgments}
	{
	This research is supported in part by Georgia Institute of Technology, IDEaS, and Microsoft Azure. S.O. was partly supported by ML@GT, Twitch, and Kwanjeong fellowships. We thank the reviewers for their feedback.
	}	

\balance 
	\bibliographystyle{ACM-Reference-Format}
	\bibliography{BIB/myref}

\end{document}